\documentclass[sigconf]{acmart}
\AtBeginDocument{%
}

\usepackage{comment}
\usepackage{xspace}
\usepackage{algorithm}
\usepackage{algpseudocode}
\usepackage{amsmath}
\usepackage{colortbl}
\usepackage{enumitem}

\copyrightyear{2026}
\acmYear{2026}
\setcopyright{cc}
\setcctype{by}
\acmConference[CIKM '26]{Proceedings of the 35th ACM International Conference on Information and Knowledge Management}{November 7--11, 2026}{Rome, Italy.}
\acmBooktitle{Proceedings of the 35th ACM International Conference on Information and Knowledge Management (CIKM '26), November 7--11, 2026, Rome, Italy}
\acmDOI{10.1145/3799682.3840082}
\acmISBN{979-8-4007-2539-5/2026/11}
\begin{document}

\title{Decomposing Staleness in Recommender Systems: A Dual-Filter Framework for Supersession and Decay}


\author{Di Bai}
\affiliation{%
\institution{Google LLC}
\city{Mountain View}
\country{USA}}
\email{vivianbai@google.com}
\orcid{0009-0009-1813-4844}

\author{Feng Han}
\affiliation{%
\institution{Google LLC}
\city{Mountain View}
\country{USA}}
\email{bladehan@google.com}
\orcid{0009-0009-5584-4562}

\author{Zhenwei Tang}
\affiliation{%
\institution{Google LLC}
\city{Toronto}
\country{Canada}}
\email{lilvjosephtang@google.com}
\orcid{0000-0002-8742-9146}

\author{Jintao Liu}
\affiliation{%
\institution{Google LLC}
\city{Mountain View}
\country{USA}}
\email{liujintao@google.com}
\orcid{0009-0005-1325-0142}

\author{Luoshu Wang}
\affiliation{%
\institution{Google LLC}
\city{Mountain View}
\country{USA}}
\email{luoshu@google.com}
\orcid{0009-0009-6681-545X}

\author{Jialu Liu}
\affiliation{%
\institution{Google LLC}
\city{New York}
\country{USA}}
\email{jialu@google.com}
\orcid{0000-0002-8721-8656}

\renewcommand{\shortauthors}{Di Bai et al.}

\begin{abstract}
Stale recommendations are a pervasive challenge and a leading source of user complaints on large-scale content platforms. Items lose relevance through two primary mechanisms: supersession, where emerging updates render prior coverage stale, and relevance decay, where an item's informational value naturally diminishes over its lifecycle. Traditional countermeasures serve as crude proxies: age cutoffs poorly reflect actual relevance loss, while engagement heuristics rely on lagging signals, broadly exposing users to stale content before the system adapts.

We present \textbf{SDF (Supersession-Decay Filtering)}, a staleness filtering system fully deployed in Google Discover, a personalized recommendation feed with hundreds of millions of daily and billions of monthly active users. SDF targets both mechanisms with complementary filters, each powered by a learned model: a \emph{relational staleness} model that detects supersession between item pairs, and a \emph{predicted traffic ratio} (PTR) model that forecasts relevance decay from the item's content, trained on lifetime visit traffic. Applied via disjunction upstream of the ranking stage, SDF prunes stale candidates, measurably reducing downstream serving costs. Online experiments demonstrate that these filters significantly reduce the prevalence of stale content while improving user engagement. Over a two-year production deployment, user-filed staleness reports (in-product user feedback) declined by 54.9\% relative to the pre-deployment baseline, establishing SDF as a robust and scalable paradigm for resolving content staleness at industrial scale.
\end{abstract}



\maketitle

\section{Introduction}
\label{sec:intro}

\begin{figure}[t]
\centering
\includegraphics[width=0.85\columnwidth]{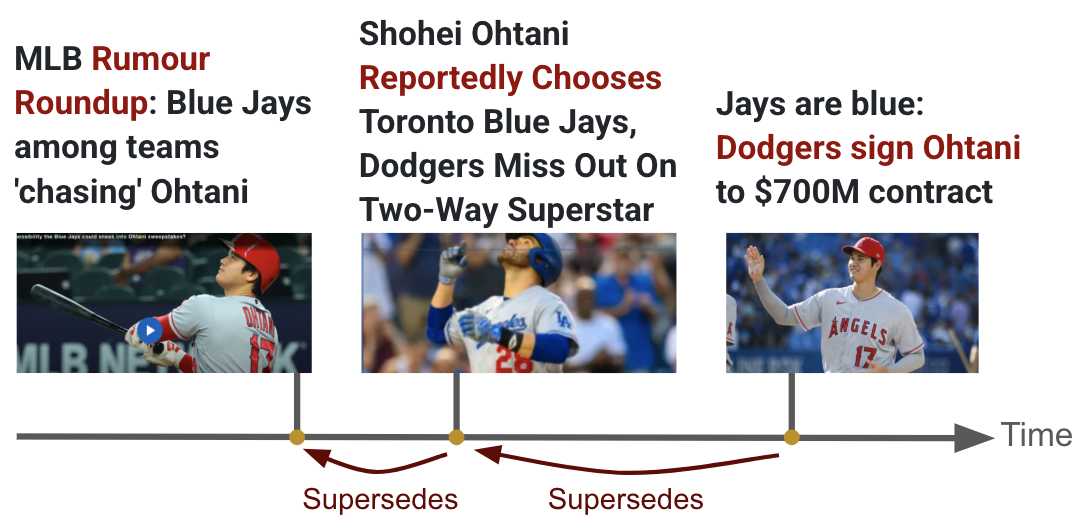}\\[3pt]
\includegraphics[width=0.85\columnwidth]{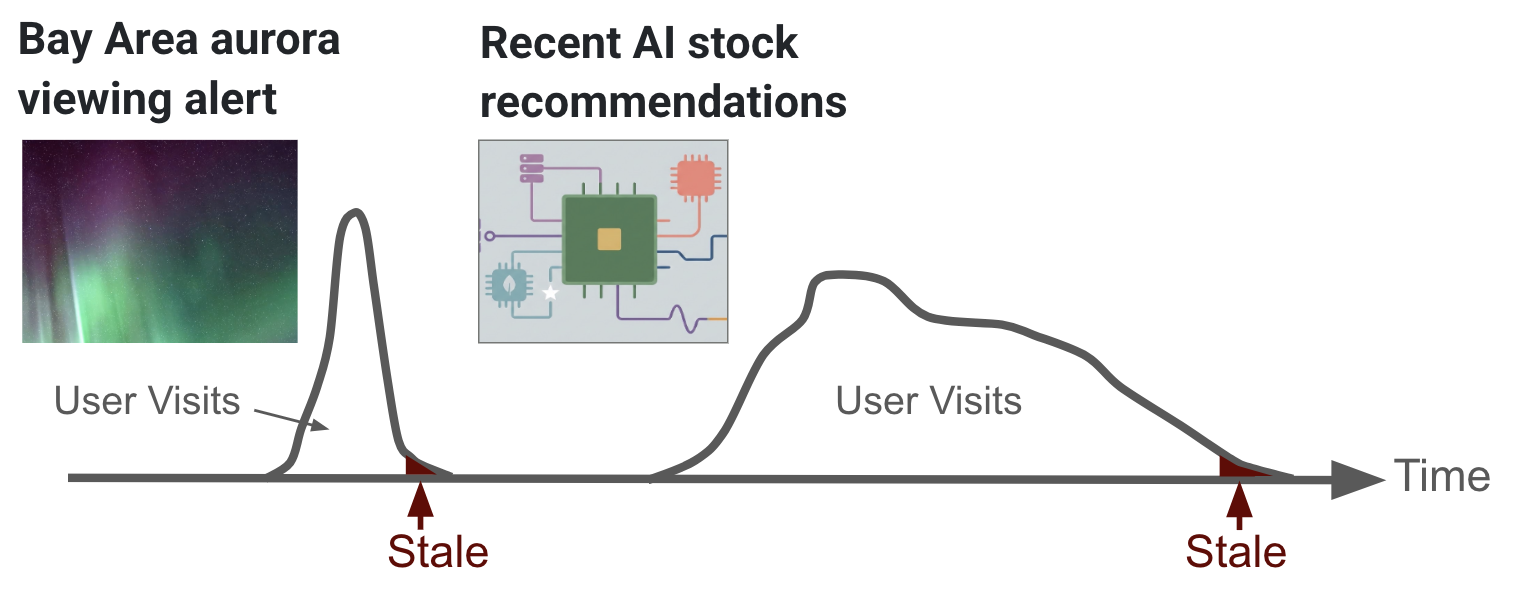}
\caption{Two staleness mechanisms. \textbf{Top: Supersession.} Each new update on the Ohtani signing renders prior coverage stale. \textbf{Bottom: Relevance decay.} Time-sensitive items naturally lose value and become stale as they near the end of their useful lifespan, without any superseding counterpart.}
\Description{Two illustrations of content staleness. The top panel is a supersession timeline in which successive news items about Shohei Ohtani's team signing are placed along a left-to-right time axis, with arrows marking that each later item renders the earlier coverage stale. The bottom panel shows user-visit-over-time curves illustrating relevance decay: a time-sensitive item spikes sharply and then goes stale within a day, while a slower-decaying item rises and falls over a longer period before going stale, each marked stale where its traffic tails off.}
\label{fig:staleness-examples}
\end{figure}

Recommender systems are inherently time-sensitive. They surface content against an evolving notion of relevance, and when this evaluation lags reality, items become stale: they continue to surface despite no longer reflecting the current state of the underlying topic. In large-scale recommender systems like Google Discover, staleness is a pervasive challenge and a primary source of user dissatisfaction. Staleness is not a monolithic problem. Items lose relevance through distinct mechanisms that dictate both how staleness manifests and how it must be detected. This work addresses two primary dimensions:

\emph{Supersession.} An item abruptly becomes stale when a newly arriving item advances the underlying topic. For example, an early bulletin reporting that ``Queen Elizabeth II is under medical supervision'' was superseded within hours by the announcement of her passing; similarly, an early report that ``Ohtani is weighing offers from multiple teams'' was rendered stale the moment his signing with the Dodgers was announced. The staleness signal here is \emph{relational} rather than absolute. Because only updates that contradict, supersede, or shift the narrative render an older item stale, detection requires comparing item pairs, not just by publication time (Figure~\ref{fig:staleness-examples}, top).

\emph{Relevance decay.} An item naturally loses informational value over its lifecycle, driven by its own content rather than supersession. A guide to ``tonight's meteor shower'' is highly useful for one evening before its relevance sharply drops, and a heartwarming local clip naturally fades without any follow-up. The staleness signal here is \emph{intrinsic} to the item, tied directly to its intended lifespan (Figure~\ref{fig:staleness-examples}, bottom).

Traditional countermeasures address neither aspect effectively. Time-based ranking demotes rather than removes stale items, forcing users to still scroll past them. Hand-tuned age cutoffs are coarse and heavily rely on publication time; a long cutoff still serves stale items, while a short one drops useful evergreen content. Engagement-based heuristics, such as click-through rate floors, introduce structural lag: the system only learns an item is stale by repeatedly exposing users to a degraded experience.

We address both mechanisms with \textbf{SDF (Supersession-Decay Filtering)}, a unified staleness filtering system deployed in Google Discover, a personalized feed of open-web content in the Google Search app, serving hundreds of millions of daily and billions of monthly active users. While downstream ranking systems remain engagement-based and trend-aware, SDF acts as a proactive filtering layer. It prunes fundamentally stale content before it enters the computation-heavy ranking stage, providing a refined candidate set. SDF accomplishes this by decomposing the problem into two complementary components:

\emph{Relational staleness filter for supersession.} To capture this relational signal, we develop a pairwise classification framework with rationale supervision that determines whether a new arrival renders an existing item stale. Because manual annotation at scale is prohibitively expensive, we leverage a large language model (LLM) pipeline for synthetic data generation and apply knowledge distillation to a student model for scalable serving.

\emph{Intrinsic staleness filter for relevance decay.} To capture intrinsic decay, we introduce the predicted traffic ratio (PTR) model. Using an item's multimodal content as input and trained on lifetime visit traffic, PTR forecasts relevance decay over future horizons. This component proactively filters items nearing the end of their lifespan, freeing serving capacity for fresher and more valuable content.

SDF combines the two filters via disjunction upstream of the ranking stage, simultaneously improving user experience and serving efficiency. Across a two-year production deployment in Google Discover, SDF reduced user-filed staleness reports (in-product user feedback) by 54.9\% (supersession by 64.0\%, decay by 34.4\%), validating the dual-filter framework for staleness filtering in recommender systems and broader content applications.
\section{Related Work}
\label{sec:related}

\smallskip\noindent\textbf{Staleness modeling in recommender systems.}
Staleness in recommender systems has primarily been studied indirectly through popularity decay rather than direct detection. One paradigm uses time as an explicit model feature (TimeSVD++~\cite{koren2009collaborative}), with later work applying personalized time-decay to click-through rate~\cite{yoneda2019algorithms} or tuning recency against relevance~\cite{chakraborty2019optimizing}. Another characterizes content lifecycles empirically: news shelf-life curves from social-media activity~\cite{castillo2014characterizing}, popularity trajectory prediction~\cite{figueiredo2014improving}, category-level lifecycle behavior~\cite{gulla2016intricacies}, and popularity dynamics~\cite{wu2007novelty, yang2011patterns, wu2023predicting}. Both lines share limitations for the staleness-filtering task: they require observed engagement, which lags, so stale content may still be served, and they model decay at coarse rather than per-item granularities. SDF's PTR is designed for this task: inspired by data-quality \emph{currency} metrics~\cite{heinrich2015metric}, it predicts content-level per-item lifetime traffic ratios from multimodal content before engagement is observed. SDF deploys PTR as an upstream filter to prune stale candidates before they consume heavy compute in downstream serving, where compute costs scale at industrial volume. A separate line, document-level novelty detection~\cite{bernstein2005redundant, soboroff2005novelty, ghosal2022novelty, zhang2002novelty}, asks whether an incoming document adds new information to a corpus; SDF's relational filter addresses the inverse problem of whether a new item renders an earlier one stale by direct pairwise relational judgment.

\smallskip\noindent\textbf{Industrial practices on staleness and corpus filtering.}
Industrial recommender systems typically combine static age cutoffs with engagement-based heuristics for candidate management, particularly to control content freshness. This pattern recurs across diverse platforms: YouTube's multi-funnel fresh-content stack~\cite{wang2023fresh} uses age and interaction-count thresholds for fresh-content slotting; Kuaishou's cold-start pipeline~\cite{chen2025cold} applies fixed exposure thresholds for cold-start gating; Twitter/X's open-source algorithm~\cite{twitter2023algorithm} applies post-ranking heuristic filters, with audits documenting significant popularity amplification~\cite{bouchaud2023crowdsourced, ye2025auditing}. Despite the recurring pattern, content-level staleness modeling remains under-explored in industry. The reliance on heuristic pipelines leaves stale items to consume substantial downstream compute, representing meaningful efficiency headroom that proactive filtering can reclaim. SDF realizes this efficiency through two complementary filters. As a self-contained module that refines the candidate set, SDF presents a generalizable framework applicable to other industrial recommender systems.

\section{Methodology}
\label{sec:method}

SDF's two filters, targeting supersession and relevance decay, are composed by disjunction. Section~\ref{sec:problem} formalizes both tasks; Sections~\ref{sec:rel-method} and~\ref{sec:ptr-method} detail each model; Section~\ref{sec:sdf-method} describes composition and serving.

\begin{figure*}[t]
\centering
\includegraphics[width=\textwidth]{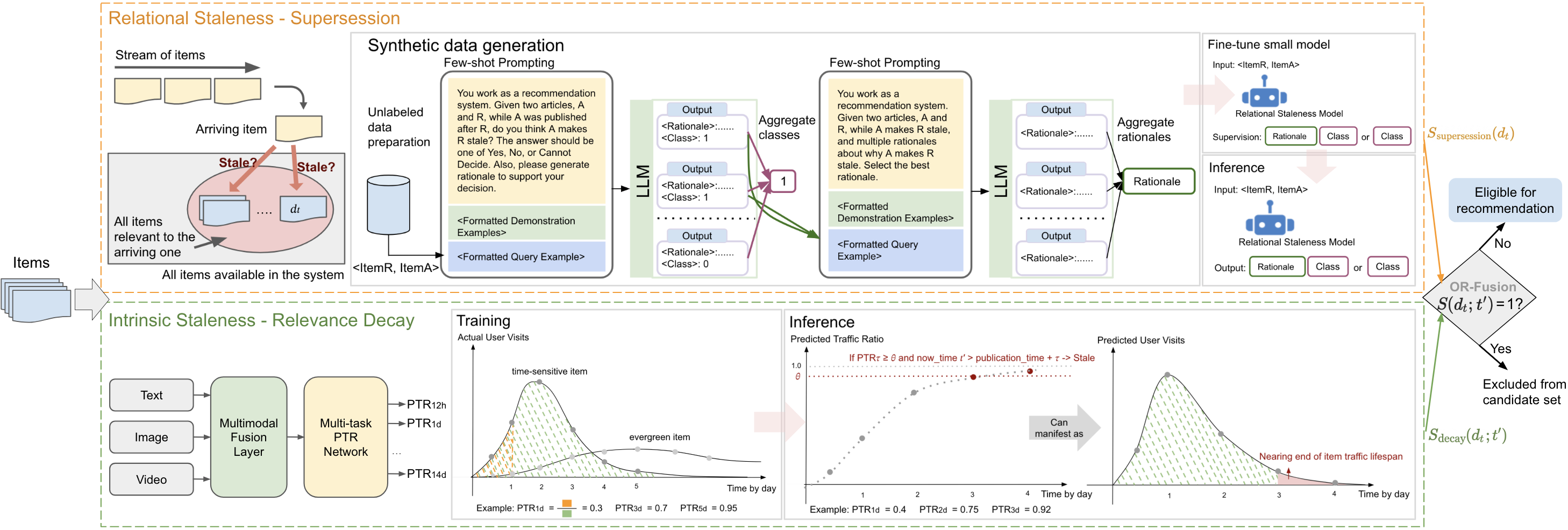}
\caption{SDF end-to-end workflow. A candidate item $d_t$ (arriving at time $t$) is evaluated in parallel. \emph{Top: Relational staleness filter for supersession.} A student model predicts pairwise staleness, yielding $S_{\mathrm{supersession}}(d_t)$. \emph{Bottom: Intrinsic staleness filter for relevance decay.} A multimodal model predicts per-horizon traffic ratios from the item's content, yielding $S_{\mathrm{decay}}(d_t; t')$ for decision time $t'$. The two outputs are OR-fused into $S(d_t; t')$; items flagged ($S(d_t; t')=1$) are filtered, while the rest remain eligible for recommendation and pass to downstream serving.}
\Description{Block diagram of the SDF end-to-end workflow with two parallel branches evaluating a candidate item. The top branch is the relational staleness filter: a student model classifies whether newer topic-related items render the candidate stale, yielding a supersession indicator. The bottom branch is the intrinsic staleness filter: the item's text, image, and video are combined in a multimodal encoder and passed to a multi-task PTR network that predicts a series of per-horizon traffic ratios; these are then combined into a decay indicator at the current decision time. The two indicators are OR-fused into a joint staleness indicator that determines whether the item stays eligible for recommendation or is excluded from the candidate set.}
\label{fig:sdf-workflow}
\end{figure*}

\subsection{Problem Formulation}
\label{sec:problem}

\smallskip\noindent\textbf{Notation.} We use the following notation:
\begin{itemize}\itemsep-1pt
\item $d$: an item in the Discover feed.
\item $\mathcal{R}(d)$: items sharing the major topic with $d$ (e.g., the same event or entity).
\item $\mathcal{A}$: the \emph{candidate set} from which the downstream ranker selects.
\end{itemize}

SDF maintains $\mathcal{A}$ continuously, adding arriving items and removing those that become stale through either of two mechanisms. The \emph{relational staleness filter} targets supersession: $d$ is removed once a newer topic-related arrival renders it stale. The \emph{intrinsic staleness filter} targets relevance decay: $d$ is removed when its content-based prediction indicates it has decayed past its useful lifespan. The downstream ranker consumes only $\mathcal{A}$, so removed items are never served.

\smallskip\noindent\textbf{Task 1: Relational staleness detection.} Learn a pairwise predictor
\begin{equation}
f(d, d') \in \{0, 1\}
\label{eq:rel-task}
\end{equation}
that takes an existing item $d$ and a newly arriving item $d' \in \mathcal{R}(d)$, and predicts whether $d'$ renders $d$ stale. Once $f(d, d') = 1$ for some $d'$, we deem $d$ stale and remove it from $\mathcal{A}$. This task addresses the \emph{supersession} aspect of staleness.

\smallskip\noindent\textbf{Task 2: Intrinsic staleness detection.} For a candidate item $d_t$ arriving in the system at time $t$ and a forward horizon $\tau$ (an elapsed time from arrival), learn a content-based predictor
\begin{equation}
g(d_t, \tau) \in [0,1]
\label{eq:int-task}
\end{equation}
that estimates the fraction of $d_t$'s expected lifetime traffic accumulating within $[t, t + \tau]$. At a later decision time $t' > t$, let $\tau = t' - t$ be $d_t$'s elapsed age. We remove $d_t$ from $\mathcal{A}$ when its predicted lifetime traffic share at this age reaches a threshold $\theta$, i.e., $g(d_t, \tau) \geq \theta$, indicating $d_t$ is predicted to have accumulated most of its lifetime traffic, with further serving yielding diminishing returns. This task addresses the \emph{relevance decay} aspect of staleness.

\smallskip\noindent\textbf{SDF fusion.} Let $\mathcal{R}_{>t}(d_t) \subseteq \mathcal{R}(d_t)$ denote items related to $d_t$ that arrived after $d_t$. We formalize the two removal rules at decision time $t'$:
\begin{equation}
S_{\mathrm{supersession}}(d_t) = \mathbf{1}\!\left[\,\exists\, d' \in \mathcal{R}_{>t}(d_t):\; f(d_t, d') = 1\,\right]
\label{eq:s-supersession}
\end{equation}
\begin{equation}
S_{\mathrm{decay}}(d_t;\,t') = \mathbf{1}\!\left[\,g(d_t, \tau) \geq \theta\,\right], \quad \tau = t' - t
\label{eq:s-decay}
\end{equation}
Their OR-fusion gives the joint staleness indicator $S(d_t;\,t')$:
\begin{equation}
S(d_t;\,t') := S_{\mathrm{supersession}}(d_t) \,\vee\, S_{\mathrm{decay}}(d_t;\,t')
\label{eq:s-fusion}
\end{equation}
The candidate $d_t$ remains in $\mathcal{A}$ when $S(d_t;\,t') = 0$. When either filter fires, $S(d_t;\,t') = 1$ and $d_t$ is removed from $\mathcal{A}$.

\subsection{Relational Staleness Detection}
\label{sec:rel-method}
To realize the pairwise predictor $f$ of Equation~\ref{eq:rel-task} under production resource constraints, we distill knowledge from an LLM teacher into a compact student model. The pipeline has two stages: synthesizing a balanced labeled pair corpus from the LLM teacher, then fine-tuning the student model on those labels (Figure~\ref{fig:sdf-workflow}).

\smallskip\noindent\textbf{Synthetic data generation.}
Supervised learning would require a large labeled dataset, but manual curation is infeasible: each pair demands careful reading of two items. We therefore synthesize the training set in two steps: selecting which item pairs to label, and obtaining labels for them from an LLM teacher~\cite{ding2023gpt,he2024annollm,tan2024large}.

\smallskip\noindent\emph{Sampling.}
Curating the unlabeled pairs is non-trivial along two axes. The first is intra-pair similarity: if it is too high, the items are likely near-duplicates with equivalent information; if too low, they are unrelated and their staleness relation is trivial. Valuable \emph{positive pairs}, where one item supersedes the other, live in a narrow mid-similarity band. The second is content redundancy: hot topics generally have more similar items, so without pre-filtering we risk generating many redundant pairs of low marginal value, even when each pair individually sits in the right similarity band. Algorithm~\ref{alg:pairs} addresses both axes: it clusters items to limit redundancy, and upweights pairs in optimal similarity ranges, so positive pairs, despite comprising a small fraction of the candidate population, are over-represented to produce a balanced dataset.

\begin{algorithm}[t]
\caption{Generate unlabeled item pairs}
\label{alg:pairs}
\small
\begin{algorithmic}[1]
\State \textbf{Input:} A set of items $\mathcal{D}$
\State \textbf{Output:} A set of item pairs $\mathcal{P}$
\State Cluster $\mathcal{D}$ and generate clusters $\mathcal{C}$
\For{cluster $c \in \mathcal{C}$}
  \State select $c^k$ diverse representative items from $c$,
  $k \propto |c|$
\EndFor
\State $\mathcal{I} \gets \bigcup_{c} c^k$
\State Similarity-score-to-weight map $M = \{[\sigma_1,\sigma_2), w_1;\,\dots\}$
\For{$d_1 \in \mathcal{I}$}
  \For{$d_2 \in \mathcal{I}$}
     \State generate pair $(d_1,d_2)$ if $d_1\neq d_2$ and from
     different clusters
     \State compute similarity $\sigma$ for the pair; look up weight
     $w \gets M(\sigma)$
  \EndFor
\EndFor
\State Generate $\mathcal{P}$ by weight-sampling over candidate pairs
\end{algorithmic}
\end{algorithm}

\smallskip\noindent\emph{Labeling.}
LLMs can adapt to tasks by learning from in-context demonstrations~\cite{brown2020language,wei2022chain}. Given an unlabeled pair $\langle d, d'\rangle$ and $K$ demonstrations $\{\langle d^{(j)}, d'^{(j)}, s_j, e_j\rangle\}_{j=1}^{K}$ for the relational staleness task, where $s_j$ is the class label and $e_j$ is a rationale, we format them into a prompt and ask the LLM to decode an output sequence containing both the class and the rationale. Mirroring conventional human annotation, where multiple raters' independent judgments are aggregated by majority vote, we draw multiple independent samples from the LLM and aggregate them by majority voting~\cite{wang2022self} (Figure~\ref{fig:sdf-workflow}). Because each annotation carries both a class and a rationale, we apply two-stage voting: first over classes, then over the candidate rationales aligned with the winning class via a second LLM pass. The second stage matters because a rationale can be wrong even when the associated class is right.

\smallskip\noindent\textbf{Model training.}
With class and rationale annotations per pair, we follow the knowledge distillation paradigm for LLMs~\cite{xu2024survey}, transferring the teacher's pairwise staleness judgments into a compact student model via supervised fine-tuning on the synthesized labels. The student consumes a \emph{seed prompt}: a fixed task prefix concatenated with the existing item $d$ and the newly arriving item $d'$. Conditioned on the seed prompt, the student decodes the rationale followed by the class in the format \texttt{``\{RATIONALE\}. Therefore, \{CLASS\}''}, patterned after Chain-of-Thought (CoT) decoding~\cite{wei2022chain}; the rationale is brief and supervised directly from the teacher~\cite{hsieh2023distilling} rather than emerging at inference time.

We additionally incorporate Natural Language Inference (NLI)~\cite{bowman2015large} into training. NLI asks a model to judge the relationship between a hypothesis and a premise, structurally analogous to judging whether an arriving item supersedes an existing one, and offers a large, human-curated corpus of such pair-relational judgments. Our synthesized staleness labels are teacher-bounded and constrained by the pair-sampling distribution of Algorithm~\ref{alg:pairs}, whereas NLI exposes the student to a broader and independently curated distribution of pair-comparison examples, strengthening its ability to judge relations between two text inputs.

\subsection{Intrinsic Staleness Detection}
\label{sec:ptr-method}

The predicted traffic ratio (PTR) model realizes $g(d_t,\tau)$ from Equation~\ref{eq:int-task} using the content of $d_t$ as input. Deployed as the intrinsic staleness filter, it complements the relational staleness filter by capturing items whose staleness arises from intrinsic decay.

\smallskip\noindent\textbf{Training target and data.}
We take $t$ to be $d_t$'s arrival time, identified with its indexing-pipeline annotation time, closely approximating its publication time. We use the user-visit distribution of an item as a natural proxy for its relevance decay: a time-sensitive item collects most of its lifetime visits within a short window after publication, while an evergreen item accrues visits more uniformly across its full lifespan. The predicted traffic ratio of an item $d_t$ at horizon $\tau$ is the share of its lifespan visit traffic that has accumulated between $t$ and $t+\tau$:
\begin{equation}
\mathrm{PTR}_\tau(d_t) =
\frac{\#\,\text{visits in }[t,\,t+\tau]}
     {\#\,\text{visits over the item's full lifespan}}
\label{eq:ptr-target}
\end{equation}
and the model is trained so that $g(d_t, \tau) \approx \mathrm{PTR}_\tau(d_t)$. Two practical specializations adapt this conceptual definition to production use in Discover. First, the denominator is operationalized as a fixed $30$-day reference window, which covers the lifespan of most content in Discover. Second, the horizon $\tau$ is sampled at a small, human-interpretable set of offsets, $\tau\!\in\!\{\text{12h, 1d, 3d, 5d, 7d, 10d, 14d}\}$, chosen to align with familiar content lifecycles: breaking news at the hour scale, daily reporting at $1$ to $3$ days, weekly content at $5$ to $7$ days, and slower-decaying pieces out to two weeks. A $\mathrm{PTR}_{7\text{d}}$ close to $1.0$ indicates the item is predicted to collect the bulk of its lifetime traffic within a week of publication and is unlikely to draw further traffic thereafter; a low value indicates an item still drawing traffic at that horizon.

PTR's training labels are derived from items' lifetime visit traffic, sourced from anonymized user activity logs. We use \emph{search-click} logs rather than raw page-view logs. Although the latter are higher-volume, they are biased by publisher promotion, making it difficult to separate time-sensitive and evergreen distributions. Search clicks instead reflect explicit, timely user intent: each click is an active user selection signaling user-perceived relevance rather than passive attention, aligning PTR's training signal with the relevance-decay phenomenon it models. PTR labels are computed by dividing the visits within each horizon by the 30-day total, yielding labels non-decreasing in $\tau$. We filter out items falling below a minimum-traffic floor to keep training labels statistically meaningful. The resulting dataset spans multiple languages and formats, giving the model coverage across Discover's serving surfaces.

\smallskip\noindent\textbf{Model architecture and filtering.}
We instantiate $g(d_t, \tau)$ as a multimodal multi-task network with early fusion (intrinsic staleness in Figure~\ref{fig:sdf-workflow}). On the input side, text, image, and video tokens of the item are combined as a single input to a shared encoder, initialized from an internal multimodal pre-trained backbone, producing a joint item representation. Horizon-specific dense feed-forward prediction heads (one per horizon) are applied to this shared representation, yielding the full set of per-horizon predictions in a single forward pass. The network is trained end-to-end with mean-squared-error loss summed across horizons.

Equation~\ref{eq:s-decay} expresses the decay decision at an idealized single horizon $\tau$ matching the item's current age $t' - t$. In practice, $g$ is trained at a fixed discrete set of horizons $\mathcal{T}$ rather than as a continuous function of $\tau$. The training labels are monotonically non-decreasing in $\tau$ by construction, and $g(d_t,\tau)$ tracks this monotonicity across $\mathcal{T}$ in practice. Equation~\ref{eq:s-decay} thus specializes to evaluating $g$ at the largest trained horizon already elapsed by decision time $t'$:

\begin{equation}
S_{\mathrm{decay}}(d_t;\,t') = \mathbf{1}\!\left[\,\max_{\tau \in \mathcal{T}} \Big(g(d_t,\tau)\cdot\mathbf{1}[\,t' > t + \tau\,]\Big) \geq \theta\,\right]
\label{eq:ptr-rule}
\end{equation}

The inner indicator masks horizons not yet elapsed; the max over the masked $g$ values isolates the largest elapsed horizon's prediction. The rule fires when that prediction meets the threshold $\theta$, i.e., the item has captured at least a fraction $\theta$ of its lifetime traffic and further serving offers diminishing value.

\subsection{SDF Composition and Serving}
\label{sec:sdf-method}
Both filters rely on unpersonalized, item-level signals: neither $f$ nor $g$ depends on the requesting user or any context other than time. They compute off the user-request path at a high enough cadence that cached outputs remain current, so applying Equation~\ref{eq:s-fusion} at serving time reduces to a lookup against the request timestamp~$t'$, with no model inference at serving time.

SDF evaluates millions of new items per day for staleness. PTR runs per-item on content alone, while the relational staleness filter runs pairwise classification, making it the compute-heavier of the two. To balance coverage with throughput, the relational staleness filter runs in two complementary modes. The \emph{online} mode, optimized for labeling latency, subscribes to incoming items and labels time-sensitive ones in near real time, using a lightweight pre-filter to drop easy negatives (e.g., unrelated pairs), plus neighbor retrieval spanning different time buckets to prevent near-duplicate items from dominating. The \emph{batch} mode, optimized for throughput, backfills the long tail and items missed by the \emph{online} mode, extending a stale label to other items in the same cluster for broader coverage.
\section{Experiments}
\label{sec:experiments}

We evaluate SDF along three axes: per-filter offline evaluation
(Section~\ref{sec:offline-eval}), online A/B testing under a
continuous prevalence-delta framework that drives launch
decisions in Discover (Section~\ref{sec:ab-test}), and the
two-year deployment outcome on user-filed reports and
serving-cost savings (Section~\ref{sec:deployment}).

\subsection{Offline Evaluation}
\label{sec:offline-eval}

\subsubsection{Relational Staleness Filter}\hfill
\label{sec:offline-eval-rel}

\smallskip\noindent\textbf{Setup.}
Following internal institutional policies, we use PaLM~2-L~\cite{anil2023palm} as the LLM annotator. We manually annotated 300 balanced pairs as a \emph{development golden set} for prompt tuning. This set is held out and disjoint from the few-shot demonstrations provided in the LLM prompt. We iteratively tuned the prompt's instructions and demonstrations until PaLM~2-L reached $>\!90\%$ precision and recall on this golden set. The tuned annotator draws 7 independent samples per pair and aggregates them by majority voting, producing labels for 1{,}100{,}133 examples (880{,}027 / 109{,}951 / 110{,}155 train/val/test). The annotator's residual error rate is on par with audited noise in widely used benchmarks, e.g., $\sim\!6\%$ on MMLU~\cite{hendrycks2020measuring,gema2025we}. We instantiate the student model of Section~\ref{sec:rel-method} as T5~\cite{raffel2020exploring}, fine-tuned from a publicly available 11B checkpoint. We train with batch size 128, learning rate $1\mathrm{e}{-3}$, 20k steps, and 1k warm-up steps. Hyper-parameters are tuned on the validation set, and metrics are reported on the test set using Accuracy (Acc.), Precision (Prec.), Recall (Rec.), and F1, with F1 as the primary metric.

\smallskip\noindent\textbf{Results.}
Table~\ref{tab:rel-eval} ablates the NLI auxiliary training and CoT-style rationale supervision. We label a configuration \emph{Disc} (discriminative) when the student is trained with class-only supervision, and \emph{Gen} (generative) when training also supervises the student on the teacher's rationale as a CoT-style signal. NLI improves F1 in both regimes, consistent with the pair-relational transfer motivated in Section~\ref{sec:rel-method}. The absolute margin is modest, but we report this ablation in full so that the design choice and its empirical support remain available to researchers and practitioners working on similar applied pipelines. Gen outperforms Disc both with and without NLI, indicating that supervising the student on the teacher's rationale at training time, rather than relying on a chain-of-thought to emerge at inference, helps independently of the auxiliary NLI training. The adopted Gen + NLI configuration achieves the best offline performance. We further validate the deployed model on a held-out 550-pair human-rated golden test set sampled from Discover with balanced positives and negatives, disjoint from the development and training sets, achieving $\mathbf{83.5\%}$ precision and $\mathbf{95\%}$ recall.

\begin{table}[t]
\centering
\caption{Relational staleness offline evaluation. Disc (discriminative): class-only supervision; Gen (generative): CoT-style rationale supervision; NLI: auxiliary pair-relational training. Per column, \textbf{bold} marks the best value and \underline{underline} the second best.}
\label{tab:rel-eval}
\begin{tabular}{l|ccc|>{\columncolor{gray!20}}c}
\toprule
 & \textbf{Acc.} & \textbf{Prec.} & \textbf{Rec.} & \textbf{F1}\\
\midrule
Disc              & 83.71 & 83.68 & 85.15 & 84.41\\
Disc + NLI        & \underline{83.86} & \underline{84.25} & 84.65 & 84.45\\
Gen               & 83.37 & 81.31 & \textbf{88.14} & \underline{84.59}\\
\midrule
Gen + NLI         & \textbf{84.16} & \textbf{84.31} & \underline{85.28} & \textbf{84.79}\\
\bottomrule
\end{tabular}
\end{table}

\smallskip\noindent\textbf{Case studies.}
Table~\ref{tab:case-study} shows CoT-style decoded outputs from the relational staleness filter, illustrating a practical advantage of the generative formulation: each prediction is accompanied by a model-decoded rationale that exposes why the arriving item renders the existing one stale, or why it does not.

\begin{table*}
\centering
\footnotesize
\begin{tabular}{p{17.6cm}}
\toprule
\textbf{Article R:} The Oakland Athletics are \colorbox{gray!20}{looking to continue their winning streak} against the struggling Texas Rangers\ldots\\
\textbf{Article A:} Rangers hope to have stabilized following series victory over Oakland. The Texas Rangers \colorbox{gray!20}{came away with a series victory}\ldots\\
\textbf{Output:} Article R is about the Oakland Athletics' goal of winning a series against the Texas Rangers. Article A reports on \colorbox{gray!20}{the outcome of that series}, making Article R outdated. Therefore, \colorbox{red!20}{Stale}.\\
\midrule
\textbf{Article R:} \colorbox{gray!20}{Fixed rate cliff} looms over Australian housing market. New CoreLogic report sheds light on emerging risk\ldots\\
\textbf{Article A:} CoreLogic has released a new report unpacking how the housing market has \colorbox{gray!20}{evolved in the three years since COVID} was declared a global pandemic.\ldots\\
\textbf{Output:} Both articles discuss the Australian housing market. Article R focuses on the fixed rate cliff, while Article A provides an overview of the market's performance over the past three years. Since \colorbox{gray!20}{Article A does not mention the fixed rate cliff}, it does not make Article R stale. Therefore, \colorbox{green!20}{Fresh}.\\
\bottomrule
\end{tabular}
\caption{CoT-style decoded outputs from the relational staleness filter. ``Article R'' is the relevant existing item; ``Article A'' is the arriving new item.}
\label{tab:case-study}
\vspace{-0.1cm}
\end{table*}

\subsubsection{Intrinsic Staleness Filter}\hfill
\label{sec:offline-eval-decay}

\smallskip\noindent\textbf{Architectural ablation: late vs.\ early fusion.}
We compare a late-fusion baseline (text, image, and video processed by separate per-modality encoders whose outputs are concatenated and fed into per-horizon dense heads) against the deployed early-fusion design of Section~\ref{sec:ptr-method} (text, image, and video tokens combined as a single input to a shared multimodal encoder before the prediction heads). Both variants are trained on identical data with matched optimizer and capacity budgets. On a curated, balanced pool of items with editorially labeled time-sensitivity, we classify each item as time-sensitive iff $g(d_t, 7\text{d}) \geq \theta$. The deployed early-fusion model reaches $\mathbf{80\%}$ accuracy, vs.\ $76\%$ for the late-fusion baseline. The four-point gap supports our early-fusion choice: jointly processing modalities outperforms per-modality late concatenation.

\begin{figure}[t]
\centering
\includegraphics[width=\columnwidth]{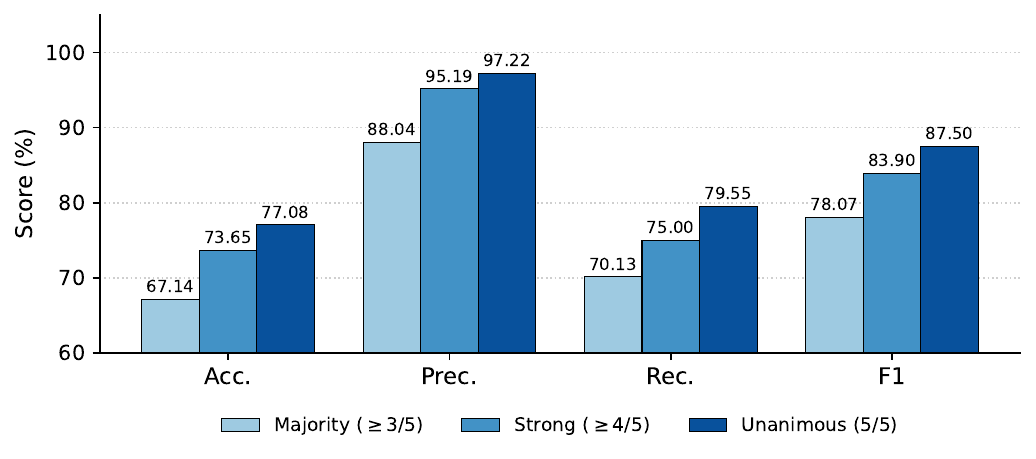}
\caption{PTR stale-item classification quality at varying rating-majority cuts; the deployed classifier ($\theta\!=\!0.9$) is held fixed. All metrics rise because tighter agreement isolates higher-confidence cases.}
\Description{Grouped bar chart of PTR stale-item classification quality, reporting accuracy, precision, recall, and F1 at three rating-agreement cuts: Majority (at least 3 of 5 raters), Strong (at least 4 of 5), and Unanimous (5 of 5). Within each metric the bars increase as agreement tightens; F1 rises from about 78 percent at the Majority cut to about 88 percent at the Unanimous cut, and precision rises from about 88 percent to about 97 percent.}
\label{fig:ptr-stale-prec}
\end{figure}

\smallskip\noindent\textbf{Stale-item classification.}
A pairwise held-out evaluation analogous to the relational staleness classifier's is not well-defined for PTR, since decay expresses itself over an item's lifetime traffic curve rather than against a fixed pair-level label. We therefore evaluate PTR directly as a stale-item classification task, matching its deployment as a binary classifier at threshold $\theta$. From a held-out pool labeled by 5 independent human raters per item, we evaluate the PTR rule of Equation~\ref{eq:ptr-rule} at $\theta\!=\!0.9$, chosen from the operating range $\{0.85, 0.9, 0.95, 0.98\}$ to keep precision high with adequate coverage. We sweep the rating-majority cut at three tiers: Majority $\geq\!3$/5, Strong $\geq\!4$/5, and Unanimous 5/5 raters agreeing whether the item is stale, with items lacking strong agreement excluded from the pool at each cut. All metrics rise as the cut tightens (Figure~\ref{fig:ptr-stale-prec}): tighter agreement shrinks the ground-truth set to the clearest cases, which the fixed classifier identifies most reliably. F1 rises from $\mathbf{78.07\%}$ at the Majority cut to $\mathbf{87.50\%}$ at the Unanimous cut; precision exceeds $\mathbf{88\%}$ across all cuts.

\subsection{Online A/B Testing}
\label{sec:ab-test}

\smallskip\noindent\textbf{Continuous prevalence-delta framework.}
User-filed staleness reports are too sparse to clear the statistical confidence bar in any single A/B test arm, so we measure online staleness by rating items sampled weekly from each A/B test's control and experiment arms across 4 consecutive weeks. Sampling is based on each item's view delta between arms, restricted to items above a minimum view delta and weighted by delta magnitude: \emph{promoted} items received more views in the experiment arm than in the control arm, while \emph{demoted} items received fewer. This differential sampling concentrates the rating budget on items the experiment arm actually moved. Each sampled item, paired with a representative view time, is rated by 5 independent human raters on an ordinal scale from \emph{Not at all stale} to \emph{Completely stale}, binarized at a fixed cutoff; an item is \emph{stale} when at least 3 of 5 raters rate it stale. Experiment metrics collapse supersession and decay staleness into a single composite metric for launch interpretability. After rating, we compute the stale rate (fraction of stale items) separately for the promoted and demoted sets. The launch metric is the \emph{prevalence delta}, the difference between these two rates:
\begin{equation}
\Delta_{\mathrm{stale}} =
r_{\mathrm{promoted}} - r_{\mathrm{demoted}}
\label{eq:prevalence-delta}
\end{equation}
where $r_X$ is the stale rate in set $X$. A negative, statistically significant $\Delta_{\mathrm{stale}}$ means stale content is more prevalent in the demoted set than the promoted set, indicating the experiment arm reduces staleness; a positive value indicates the reverse. Per-week ratios are averaged for launch reporting.

\smallskip\noindent\textbf{Filter A/B results.}
Table~\ref{tab:prevalence-delta} compares SDF and its two component filters individually alongside a baseline of two heuristics commonly used in industry, age cutoffs and an engagement cliff, in the initial locale experiment. Configurations are tuned to be engagement-neutral; trivially aggressive cuts could reduce staleness much further but at the cost of substantially shrinking the candidate set and degrading user engagement, which we explicitly rule out. Within this engagement-neutral constraint, each SDF filter individually drives a statistically significant negative $\Delta_{\mathrm{stale}}$, while the full SDF achieves the strongest reduction. This indicates that the two filters target largely distinct staleness mechanisms and work well together.

\begin{table}[t]
\centering
\caption{Prevalence delta of SDF and its component filters individually, alongside an age-cutoff + engagement-cliff baseline. Negative $\Delta_{\mathrm{stale}}$ means the promoted set is less stale than the demoted set. For $\Delta_{\mathrm{stale}}$, \textbf{bold} marks the best result.}
\label{tab:prevalence-delta}
\begin{tabular}{l|cc|>{\columncolor{gray!20}}c}
\toprule
\textbf{Filter} & \textbf{Promoted} & \textbf{Demoted}
& \textbf{$\Delta_{\mathrm{stale}}$}\\
\midrule
Baseline        & $4.00 \pm 0.86\%$ & $4.10 \pm 0.87\%$
                & $-0.10 \pm 1.22\%$\\
Relational      & $1.68 \pm 0.81\%$ & $5.87 \pm 2.11\%$
                & $-4.19 \pm 2.25\%$\\
Intrinsic       & $7.90 \pm 1.67\%$ & $11.00 \pm 1.94\%$
                & $-3.10 \pm 2.56\%$\\
\midrule
SDF             & $4.33 \pm 0.89\%$ & $11.24 \pm 2.04\%$
                & $\mathbf{-6.91} \pm 2.23\%$\\
\bottomrule
\end{tabular}
\end{table}

\smallskip\noindent\textbf{Long-term joint-holdback experiment.}
A two-month joint-holdback experiment, in which both
SDF filters are disabled for a held-out user
population, quantifies the user-side impact alongside the
staleness movement. Compared to
the holdback, SDF drives a
$\mathbf{-0.16} \pm 0.07\%$ reduction in user dismissal rate, a
$\mathbf{+0.13} \pm 0.04\%$ lift in feed diversity, and a
$\mathbf{+0.26} \pm 0.18\%$ lift in feed positive engagement; all $\pm$ ranges report 95\% confidence intervals.
The pattern is consistent with long-term system health: users
dismiss fewer items, feed diversity ticks up, and
positive engagement lifts, despite the reduction in candidate volume from filtering.

\subsection{Online Deployment}
\label{sec:deployment}

\begin{figure}[t]
\centering
\includegraphics[width=\columnwidth]{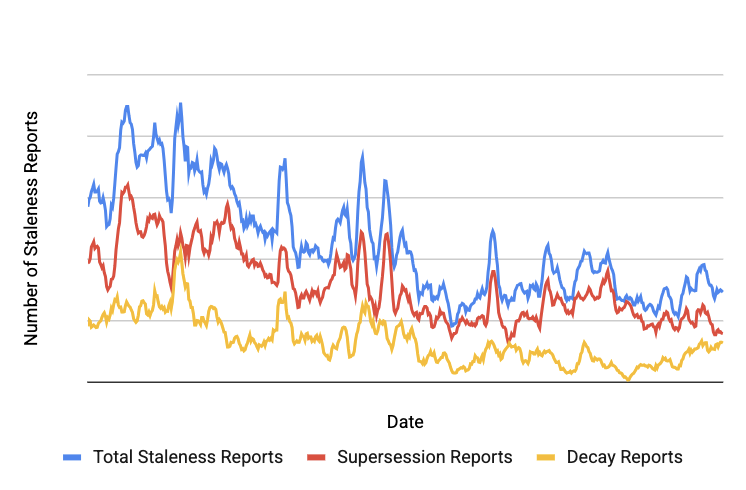}\\[2pt]
\includegraphics[width=\columnwidth]{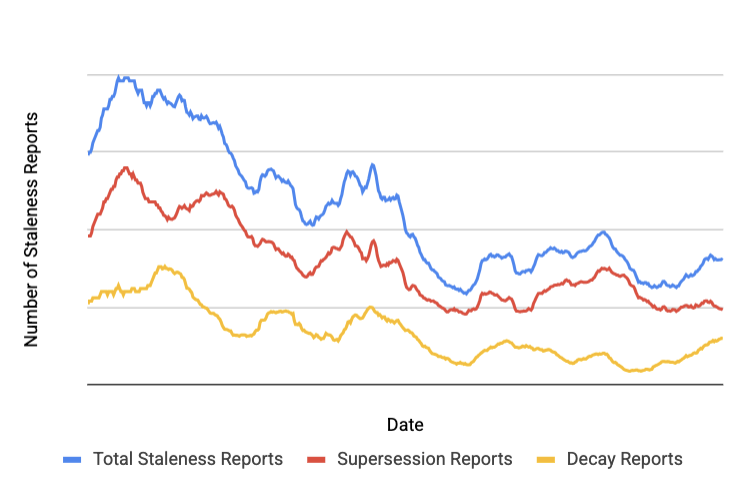}
\caption{User-filed staleness reports over the two-year SDF
deployment in Discover, by mechanism: total (blue),
supersession (red), decay (yellow).
\textbf{Top:} 7-day rolling average. \textbf{Bottom:} 28-day rolling
average.}
\Description{Two line charts of user-filed staleness report volumes over the two-year SDF deployment, each plotting three curves: total reports (blue), supersession reports (red), and decay reports (yellow), against time on the horizontal axis. The top chart shows the 7-day rolling average and the bottom chart the 28-day rolling average. All three curves trend downward over the window, with the total and supersession curves declining most, indicating fewer staleness reports after deployment.}
\label{fig:reports-longitudinal}
\vspace{-0.3cm}
\end{figure}

\smallskip\noindent\textbf{User-reported staleness over a 2-year
window.}
We monitor user-filed staleness reports over a two-year window of SDF deployment, covering successive model upgrades and locale expansion, with a stable user base (Figure~\ref{fig:reports-longitudinal}). Reports are filed by selecting \emph{stale} from a quality-issue menu. Each is then categorized as supersession or decay. We track both 7-day and 28-day rolling averages; the 28-day view is the official monitoring metric, smoothing weekly and event-driven variance. Compared to the start-of-window pre-deployment baseline, total reports fell by $\mathbf{54.9\%}$, supersession by $\mathbf{64.0\%}$, and decay by $\mathbf{34.4\%}$, each primarily driven by its corresponding filter, consistent with SDF's architectural decomposition.

\smallskip\noindent\textbf{Wins and losses.}
We present representative production cases that illustrate SDF's wins and losses.
\begin{itemize}[noitemsep,topsep=0pt,leftmargin=*]
\item \emph{Wins.} The relational staleness filter correctly flags pairwise news supersession: \emph{``House GOP leaders float new plan''} (existing) is flagged stale by \emph{``House Republicans failed to agree on a spending plan''} (arriving), and \emph{``Mazda 3 IPM launching soon?''} (existing) by \emph{``Mazda 3 prices revealed''} (arriving). PTR correctly drops single-day items (Geminid meteor-shower guide) after their event window, retains multi-day items (Phuket food-festival guide) during their relevance period, and leaves evergreen items (NYT health-coach explainer) unfiltered.
\item \emph{Losses.} (i) The relational staleness filter's pairwise judgment can be ambiguous on certain items such as political commentary, where content tends to mix opinion and fact and no single item cleanly supersedes another. (ii) PTR misses a long tail of gradually-decaying items: their predicted traffic ratio never reaches the threshold, and some become moderately stale without being caught. (iii) A substantial fraction of residual user-filed staleness reports corresponds to content the user has already seen on Discover or elsewhere. Such user-perceived staleness lies outside SDF's supersession/decay scope.
\end{itemize}

\smallskip\noindent\textbf{Serving-cost savings.}
SDF prunes candidates, avoiding heavy downstream workloads such as dense scoring. Monitored over Q1 2026, the relational staleness filter saves $\mathbf{4.76\%}$ in serving CPU and TPU hours, while the intrinsic staleness filter saves an additional $\mathbf{3.98\%}$. Across a continuously updating corpus, these efficiency gains translate directly into proportional cost reductions.

\smallskip\noindent\textbf{Continuing development.}
SDF has been deployed across a two-year window marked by rapid LLM advances. We have been progressively upgrading the relational staleness model with newer LLM releases; closing the gap to the latest LLM capabilities within serving's latency and cost budgets remains an ongoing effort. In parallel, we are actively augmenting PTR with multimodal LLM reasoning and extending it toward continuous-horizon lifespan representation learning.

\section{Conclusion}
\label{sec:conclusion}
Recommender systems contend with continuously shifting relevance. We present SDF, which reframes staleness as a joint property of an item and its environment, with two complementary dimensions: a \emph{relational} dimension capturing supersession by newer items, and an \emph{intrinsic} dimension capturing relevance decay over the item's own lifecycle. We address the relational dimension with an LLM-distilled pairwise model and the intrinsic dimension with a multimodal predicted traffic ratio model, composing the two via OR-fusion to check content-serving eligibility. Across a two-year deployment in Google Discover, a personalized recommendation feed serving hundreds of millions of daily and billions of monthly active users, SDF reduced user-filed staleness reports by $54.9\%$, eased user dissatisfaction, and enhanced positive engagement. Beyond Discover, SDF establishes the supersession-decay paradigm as a principled, scalable, and widely applicable approach to staleness in recommender systems and broader content applications.


\section*{GenAI Usage Disclosure}
The authors acknowledge the use of Google Gemini to assist with the refinement of the manuscript. All AI-assisted revisions were manually reviewed by the authors, who assume full responsibility for the accuracy and integrity of the final work.

\bibliographystyle{ACM-Reference-Format}
\bibliography{references}


\end{document}